\documentclass{emulateapj}
\usepackage{ulem}
\usepackage{longtable}
\usepackage{graphicx}
\input{epsf.sty}
\input{psfig.sty}

\shorttitle{BH spin in Sw J1644+57 and Sw J2058+05}

\begin{document}
\title{Black hole Spin in Sw J1644+57 and Sw J2058+05}

\author{Wei-Hua Lei\altaffilmark{1,2}, Bing Zhang\altaffilmark{1}}

\altaffiltext{1}{Department of Physics and Astronomy, University of Nevada Las Vegas, 4505 Maryland Parkway, Box 454002, Las Vegas, NV 89154-4002, USA. Email: leiwh@physics.unlv.edu; zhang@physics.unlv.edu}

\altaffiltext{2}{School of Physics, Huazhong University of Science and
Technology, Wuhan, 430074, China}

\begin{abstract}
Recently a hard X-ray transient event, Sw J1644+57, was discovered by the Swift satellite. It likely marks the onset of a relativistic jet from a supermassive black hole, possibly triggered by a tidal disruption event. Another candidate
in the same category, Sw J2058+05, was also reported. The low event rate
suggests that only a small fraction of TDEs 
launch relativistic jets. A common speculation is that these
rare events are related to rapidly spinning black holes. We attribute
jet launching to the Blandford-Znajek mechanism, and use the available
data to constrain the black hole spin parameter for the two events. It is
found that the two black holes indeed carry a moderate to high spin,
suggesting that black hole spin is likely the crucial factor behind the
Sw J1644+57 - like events.
\end{abstract}

\keywords{ accretion, accretion disks--black hole physics--magnetic fields}

\section{Introduction}
A hard X-ray transient event, Swift J16449.3+573451 (``Sw J1644+57'' 
hereafter) was discovered by the Swift satellite (Gehrels et al. 2004)
on 2011 March 28, initially reported as GRB 110328A
(Cummings et al. 2011). It was soon realized that it is not a regular
GRB. Long-term follow up observations with Swift (Burrows et al. 2011)
revealed that it has extended emission in the X-ray band without
significant decay. A much longer variability time scale $\delta t \sim
100$ s than normal GRBs (Burrows et al. 2011) 
as well as its location near the center of a $z=0.354$
host galaxy (Levan et al. 2011) link the source to a 
super-massive black hole (BH) with $M_{\bullet}\sim 10^7 M_{\sun}$
(Burrows et al. 2011; Bloom et al. 2011). The super-Eddington X-ray
luminosity (Burrows et al. 2011), bright radio afterglow 
(Zauderer et al. 2011), as well as a historical stringent X-ray flux 
upper limit suggest that the event marks the onset of a relativistic 
jet from a super-massive BH. The sharp onset, and gradual fade-away
of X-ray flux suggest that the transient may be triggered by 
tidal disruption of a star by the BH (Bloom et al. 2011;
Burrows et al. 2011).
 
A second candidate in the same category, Swift J2058.4+0516 
(``Sw J2058+05'' hereafter) was reported by Cenko et al. (2011),
who argued that its observational properties are rather similar to 
those of Sw J1644+57. 

These events are rare (2 detected in 7 years of Swift operation). Most previous studies of tidal disruption events (TDEs) from super-massive BHs did not expect an associated relativistic jet (e.g. Kobayashi et al. 2004; Strubble \& Quataert 2010; Lodato \& 
Rossi 2011, but see Lu et al. 2008; Giannios \& Metzger 2011).
The neutrino-annihilation mechanism as invoked in GRBs is 
inadequate given a much lower accretion rate as compared with
GRBs (e.g. Shao et al. 2011). A plausible mechanism to launch a jet
from such an event is to tap spin energy of the BH through a magnetic 
field, which connects the BH event horizon and a remote astrophysical 
load (Blandford \& Znajek 1977, here after BZ). This mechanism has
been widely invoked to interpret AGN jets (e.g. Vlahakis \& Konigl 
2004). Within this scenario, a dominant magnetic jet composition 
is envisaged. Indeed, modeling the emission of Sw J1644+57 suggests 
that the jet is highly ``particle starved'', i.e. most energy is 
not carried by particles. A natural inference is that the outflow
is Poynting-flux-dominated (Burrows et al. 2011).

If one identifies the BZ mechanism as the jet launching mechanism,
the BH spin parameter (a dimensionless angular momentum of the BH, which ranges from 0 for no spin to 1 for the maximum spin) can be constrained from the data. This is the subject in this {\it Letter}.

\section{Method}

The bipolar BZ jet power from a BH with mass $M_{\bullet}$ and 
angular momentum $J$ is (Lee et al. 2000; Li 2000; Wang et al. 2002;
McKinney 2005) 
\begin{equation}
L_{\rm BZ}=1.7 \times 10^{44} a_{\bullet}^2 M_{\bullet,6}^2 
B_{\bullet,6}^2 F(a_{\bullet}) \ {\rm erg \ s^{-1}},
\label{eq_Lmag}
\end{equation}
where $a_{\bullet}=Jc/(G M_{\bullet}^2)$ is the BH spin parameter, 
$M_{\bullet,6}=M_{\bullet}/10^6M_{\sun}$, 
$B_{\bullet,6}=B_{\bullet}/10^6 {\rm G}$ and
\begin{equation}
F(a_{\bullet})=[(1+q^2)/q^2][(q+1/q) \arctan q-1]
\label{eq_F}
\end{equation}
here $q= a_{\bullet} /(1+\sqrt{1-a^2_{\bullet}})$, and $2/3\leq F(a_{\bullet}) \leq \pi-2$ for 
$0\leq a_{\bullet} \leq 1$. It apparently depends on $M_{\bullet}$, $B_{\bullet}$,
and $a_{\bullet}$. However, since an isolated BH does not carry a magnetic field,
$B_{\bullet}$ is closely related to the accretion rate $\dot M$ and the
radius of the BH, which depends on $M_{\bullet}$. Combining these 
dependences, one finds that $L_{\rm BZ}$ is essentially independent
of $M_{\bullet}$, but is rather a function of $\dot M$ and $a_{\bullet}$. 
This may be proven with the following rough scalings.
For a Newtonian disk, angular momentum equation states 
$\dot{M}r^2 (GM_{\bullet}/r^3)^{1/2} \simeq - 4 \pi r^2 \tau_{r\phi} h$, 
where $\tau_{r\phi}$ is the viscous shear, and $h \propto r$ is the half 
disk thickness, with $h/r \ll 1$ for a thin disk and 
$h/r \sim 1$ for a thick disk. Adpoting the $\alpha$-prescription 
for viscosity, the viscous shear can be expressed as $\tau_{r\phi} 
=-\alpha P$, where $P$ is the total pressure in the disk. 
One therefore derives $P \simeq \dot{M} r^2 (GM_{\bullet}/r^3)^{1/2}/(4\pi 
r^2 \alpha h) \propto (M_{\bullet}/r^5)^{1/2} \propto 
M_{\bullet}^{-2}$, where we have applied the scaling
$r \propto r_s \propto M_{\bullet}$ ($r_s = 2 G M_{\bullet}/c^2$ is the
Schwarzschild radius). To make an efficient BZ jet, the accretion
inflow should carry a large magnetic flux (e.g. Tchekhovskoy et al.
2011). It is reasonable to assume that magnetic fields in the disk
are in close equilibrium with the total pressure, so that 
$B_{\bullet}^2 \propto P \propto M_{\bullet}^{-2}$. Inserting this
dependence to Eq.(\ref{eq_Lmag}), one finds $L_{\rm BZ}$ is
essentially independent of $M_{\bullet}$.

More precisely, we adopt the following prescription to treat
the problem. The total pressure in the disk can be expressed 
as $P = P_{\rm rad}+P_{\rm gas}+P_{\rm B}$, where $P_{\rm rad}
=a T^4/3$, $P_{\rm gas}=n k T$, and $P_{\rm B}=B_{\rm disk}^2/8 \pi$ 
are radiation, gas and magnetic pressure, respectively. 
Here $T$ is the temperature of the disk, $n$ is the gas
particle number density, $a$ is radiation density constant, 
and $k$ is Boltzmann constant. We denote $P_{\rm B} = \beta
P$, and take $\beta \sim 0.5$ in this work. This corresponds
to a maximized magnetic flux. A smaller $\beta$ would demand
an even larger $a_{\bullet}$ to reach a same BZ power. So taking
$\beta \sim 0.5$ gives an estimate of a conservative lower 
limit of $a_{\bullet}$. For the two sources (Sw J1644+57 and 
Sw J2058+05) we are interested in, the accretion rate is estimated 
close to or even larger than the Eddington accretion rate (Burrows 
et al. 2011; Cenko et al. 2011). In this regime, a thin disk model 
may be adequate to describe the disk, and we apply it to estimate 
disk pressure for simplicity. The disk pressure peaks in the inner 
region where radiation pressure may dominate. As a result, for 
$\beta\sim 0.5$, one has
\begin{equation}
\frac{B_{\rm disk}^2}{8\pi} \sim \frac{1}{3} a T^4~,
\label{eq_B}
\end{equation}
where the temperature of the disk ($a_\bullet$- and $r$-dependent) 
can be written
\begin{equation}
T(a_\bullet,r) = \left(\frac{3G M \dot{M}}{8\pi r^3 \sigma} f\right)^{1/4}
\label{eq_T}
\end{equation}
where $\sigma$ is the Stephan-Boltzmann constant, and
the general relativistic correction factor (Page \& Thorne 1974)
\begin{eqnarray}
f(a_\bullet,r) & = & \frac{\chi^2}{(\chi^3-3\chi+2a_{\bullet})} \left[ \chi-\chi_{\rm in} -\frac{3}{2}a_{\bullet} \ln\left(\frac{\chi}{\chi_{\rm in}}\right)- \right. \nonumber \\
&  & \frac{3(\chi_1-a_{\bullet})^2}{\chi_1(\chi_1-\chi_2)(\chi_1-\chi_3)} \ln\left(\frac{\chi-\chi_1}{\chi_{\rm in}-\chi_1}\right) -  \nonumber \\
&  & \frac{3(\chi_2-a_{\bullet})^2}{\chi_2(\chi_2-\chi_1)(\chi_2-\chi_3)} \ln\left(\frac{\chi-\chi_2}{\chi_{\rm in}-\chi_2}\right)- \nonumber \\
&  & \left. \frac{3(\chi_3-a_{\bullet})^2}{\chi_3(\chi_3- \chi_1)(\chi_3-\chi_2)} \ln\left(\frac{\chi-\chi_3}{\chi_{\rm in}-\chi_3}\right)  \right]~,
\label{eq_fgr}
\end{eqnarray}
where $\chi=(r/r_g)^{1/2}$, $\chi_{\rm in}=(r_{\rm in}/r_g)^{1/2}$, 
and $r_g=GM_{\bullet}/c^2$. For a Kerr BH, the disk inner edge 
$r_{\rm in}$ is expressed as (Bardeen et al. 1972),
\begin{eqnarray}
r_{\rm in}/r_g =  3+Z_2 -\left[(3-Z_1)(3+Z_1+2Z_2)\right]^{1/2},
\end{eqnarray}
for $0\leq a_{\bullet} \leq 1$, where $Z_1 \equiv 1+(1-a_{\bullet}^2)^{1/3} [(1+a_{\bullet})^{1/3}+(1-a_{\bullet})^{1/3}]$, $Z_2\equiv (3a_{\bullet}^2+Z_1^2)^{1/2}$. 
In Eq.(\ref{eq_fgr}), $\chi_1,\chi_2,\chi_3$ are the three roots of 
$\chi^3 - 3\chi+ 2 a_{\bullet} = 0$, i.e. $\chi_1= 2 \cos(\frac{1}{3} \cos^{-1}a_{\bullet} - \pi/3)$, $\chi_2= 2 \cos(\frac{1}{3} \cos^{-1}a_{\bullet} + \pi/3)$, $\chi_3=- 2 \cos(\frac{1}{3} \cos^{-1}a_{\bullet} )$.
It is easy to check that $f(r=r_{\rm in})=0$ and $f(r \gg r_{\rm in})\simeq 1$. For a Newonian disk, $f$ can be simply written as $f = 1-\sqrt{r_{\rm in}/r}$.

We assume that the strength of the magnetic field threading the BH is
comparable to the largest field strength in the disk, i.e.
$B_\bullet \sim B_{\rm disk}^{\rm max}$. For given $\dot M$ and 
$a_\bullet$, we solve numerically the temperature profile of the thin 
disk and identify the radius $r_{\rm peak}$ where $T$ reaches the 
maximum. It is found that $r_{\rm peak}$ is very close to
$r_{\rm in}$, the inner most radius of the accretion disk. 
We then calculate $B_{\rm disk}^{\rm max}$ from Eq.(\ref{eq_B}), 
which is assigned to $B_\bullet$. Applying Eqs.(\ref{eq_Lmag})
and (\ref{eq_F}), one then obtains $L_{\rm BZ}$ once $\dot M$
and $a_{\bullet}$ are specified.

Observationally, a time-dependent isotropic X-ray luminosity 
$L_{\rm X,iso}$ was measured for the two sources. This luminosity 
can be connected to the BZ power through
\begin{equation}
\eta L_{\rm BZ}=f_b L_{\rm X,iso}
\label{eq1}
\end{equation}
where $\eta$ is the efficiency of converting BZ power to X-ray radiation,
\begin{equation}
f_b = \frac{\Delta \Omega}{4\pi} = {\rm max} 
\left(\frac{1}{2\Gamma^2}, \frac{\theta_j^2}{2}\right)< 1
\end{equation} 
is the beaming factor of the jet (Burrows et al. 2011), 
$\Delta\Omega$ is the solid angle of the bipolar jet, and
$\Gamma$ and $\theta_j$ are the Lorentz factor and opening angle
of the jet.

In this work, we adopt $\eta \sim 0.5$, motivated by a potentially
high radiation efficiency of a magnetically-dominated jet 
(e.g. Drenkhahn \& Spruit 2002; Zhang \& Yan 2011). Again this gives 
a conservative lower limit of the inferred $a_{\bullet}$, since a less 
efficient jet would demands an even higher spin rate in order to 
interpret the same observed luminosity.

The parameter $f_b$ has a large uncertainty. Based on the current 
data, one cannot precisely measure $\Gamma$ and $\theta_j$. Here we
apply a statistical method to estimate the range of $f_b$ within
the TDE framework (see also Burrows et al. 2011). First, the facts 
that two such events (Sw J1644+57 and Sw J2058+05) were detected by 
Swift in $\sim 7$ years and that the field view of Swift Burst Alert
Telescope (BAT, Barthelmy et al. 2005) is $\sim 4\pi/7$ sr suggest that
the all-sky rate of such events is $R_{\rm obs} \sim 2 {\rm yr}^{-1}$, 
with a 90\% confidence interval of $(0.44 - 5.48) {\rm yr}^{-1}$ 
(Kraft et al. 1991). Next, the TDE rate is estimated as 
$\sim 10^{-5} - 10^{-4} {\rm yr^{-1}~ galaxy^{-1}}$ based on
observational (Donley et al. 2002; Gezari et al. 2009) theoretical 
(Wang \& Merritt 2004) studies. The galaxy number density is $n_{\rm gal} 
\sim 10^{-3}-10^{-2} {\rm Mpc}^{-3}$ (Tundo et al. 2007). Sw J2058+05
was marginally detected at $z=1.1853$ (Cenko et al. 2011). We then 
obtain the total TDE event rate within the volume ($z\leq 1.1853$) 
$R_{\rm tot} \sim 10^4 -10^5 {\rm yr}^{-1}$. Considering that only $\sim 10\%$
of the population can launch a jet (the ``radio-loud'' AGN fraction,
Kellerman et al.1989; Cirasuolo et al. 2003), the beaming factor 
can be estimated as
\begin{equation}
f_b \sim \frac{R_{\rm obs}}{10\% R_{\rm tot}} \in (4.4\times 10^{-5},
5.5 \times 10^{-3})~.
\end{equation}

Finally, in order to infer $a_{\bullet}$ from $L_{\rm X,iso}$, $f_b$ (given
$\beta=0.5$ and $\eta=0.5$), one needs to know the accretion rate
$\dot M$. This is an even loosely constrained parameter. However,
if one assumes that the luminosity history of the light curve
well delineates the accretion history of the BH (noticing 
$L_{\rm BZ} \propto \dot M$), one can normalize
the accretion rate using the total accreted mass based on the
observed flux and fluence of the event. For Sw J1644+57, since the 
source has entered a decaying phase, during which the residual fluence 
no longer contributes significantly to the total fluence, one can 
take the current total X-ray fluence as a good proxy of the total 
mass of the tidally disrupted star. Taking the peak flux as example,
the peak accretion rate can be estimated as
\begin{equation}
\dot M_{\rm peak} = \frac{F_{\rm X}^{\rm peak}(1+z)}{S_{\rm X}} M_*~
=\frac{L_{\rm X,iso}^{\rm peak}}{E_{\rm X,iso}} M_*,
\label{Eq_Mdot}
\end{equation}
where $F_{\rm X}^{\rm peak}$ is the peak X-ray flux,  $S_{\rm X}$ is the 
total X-ray fluence, $L_{\rm X,iso}^{\rm peak}$ is the peak isotropic 
X-ray luminosity, $E_{\rm X,iso}$ is the isotropic X-ray energy,
and $M_*$ is the total mass of the star. The factor $(1+z)$ was
applied to convert the observed time to the time in the source
rest frame. The accretion
rate at other epochs can be estimated similarly. The range of
$M_*$ may be between $0.1 M_\odot$ and $10 M_\odot$. One can then
derive a mass-dependent constraint on $a_{\bullet}$.

\section{Sw J1644+57 and Sw J2058+05}

Now we apply the above method to the two sources.

According to Burrows et al. (2011), during the first 50 days after 
the first BAT trigger the total
X-ray energy corrected for live-time fraction for Sw J1644+57
is $E_{\rm X,iso}({\rm J1644}) \sim 2 \times 10^{53}$ erg in the
1.35-13.5 keV rest-frame energy band (corresponding the energy band
of Swift X-Ray Telescope, Burrows et al. 2005). The peak 
luminosity in the same energy band is $L_{\rm X,iso}^{\rm peak} 
({\rm J1644}) \sim 2.9 \times 10^{48}~{\rm erg~s^{-1}}$. 
The accretion rate (Eq.(\ref{Eq_Mdot})) can be estimated as
\begin{equation}
\dot{M}_{\rm peak} ({\rm J1644}) \simeq 1.45 \times 10^{-5} M_* {\rm s}^{-1}~. 
\label{eq_dotMp}
\end{equation}

\begin{figure}[htc]
\center
\includegraphics[width=7cm,angle=0]{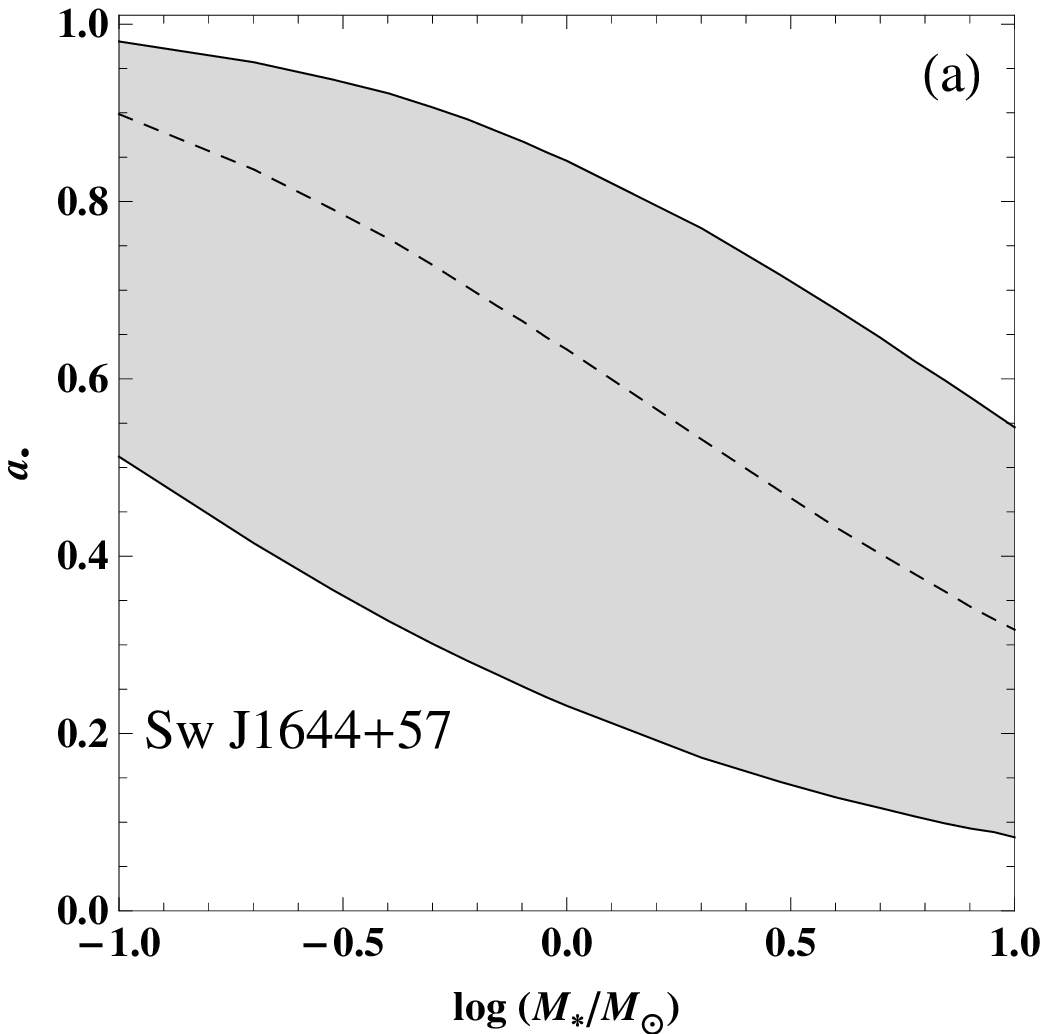}
\includegraphics[width=7cm,angle=0]{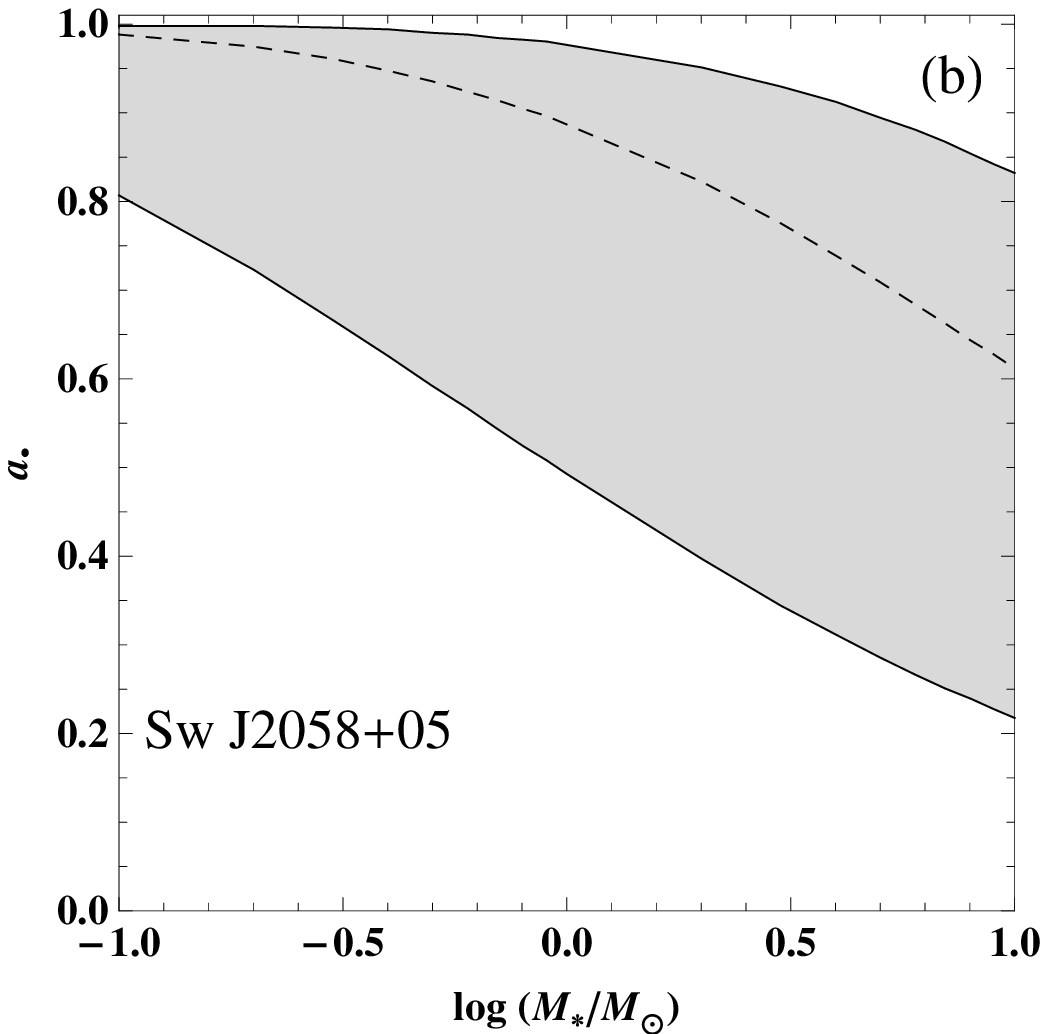}
\caption{Parameter space of $M_* -  a_{\bullet}$ for Sw J1644+57 (left panel) and Sw J2058+05 (right panel) with $\beta =0.5$ and $\eta=0.5$. The shaded region indicate the range for BH spin bracketed by $f_b = 4.4\times 10^{-5}$ (lower) and $f_b = 5.5 \times 10^{-3}$ (upper). The dashed line corresponds to the most probable value $f_b=10^{-3}$. For Sw J2058+05, $\zeta = 0.1$ is adopted. It is shown that the most probable values are $a_{\bullet} ({\rm J1644}) = 0.63, 0.90$ and $a_{\bullet} ({\rm J2058}) = 0.89, 0.99$ for $M_* = 1, 0.1 M_\odot$, respectively. }
\label{fig_Ma}
\end{figure}

Fig.\ref{fig_Ma}a presents the constraint on $a_{\bullet}$ for Sw J1644+57
as a function of $M_*$, with $\beta=0.5$, $\eta=0.5$. The two boundary
lines correspond to two ends of the range of $f_b$, with the lower and
upper boundaries correspond to $f_b=4.4\times 10^{-5}$ and $f_b=5.5
\times 10^{-3}$, respectively. The middle dashed line corresponds to the 
most probable value $f_b \sim 10^{-3}$. One can see that the supermassive
BH is demanded to have a moderate to high spin rate. Given the standard
stellar initial mass function, the number of low-mass stars is much 
more abundant than the high-mass stars. If one takes $M_* = 1 M_\odot$,
the required range of $a_{\bullet}$ is $(0.23, 0.85)$, with the most probably value
$a_{\bullet} = 0.63$. For $M_* = 0.1 M_\odot$ (more probable), the range of $a_{\bullet}$
is $(0.51, 0.98)$ with the most probable value $a_{\bullet} = 0.90$.

Sw J2058+05 was discovered by Swift BAT through a 4-day (2011 May 17-20)
integration. A subsequent target-of-opportunity (ToO) observation 8 days 
after the end of 4-day integration (2011 May 28) revealed an X-ray source 
that behaves very similarly to Sw J1644+57 (Cenko et al. 2011), suggesting
that it is very likely another Sw J1644+57-like event. The $0.3-10$ keV 
peak flux is $F_{\rm X,iso}^{\rm peak} \simeq 7.9 \times 10^{-11} {\rm erg ~ 
cm^{-2} ~ s^{-1}}$, corresponding to a peak luminosity
$L_{\rm X,iso}^{\rm peak} \simeq 3\times 10^{47}~{\rm erg~s^{-1}}$. 
The total X-ray fluence from the beginning of the
ToO observation to 2011 July 20 is $S_{\rm X} \simeq 1.0 \times 10^{-4} 
{\rm erg ~ cm^{-2}}$ (Cenko et al. 2011). Since we did not catch the
X-ray emission during the first 11 days (May 17 - 27) when it is supposed
to be much brighter, the registered X-ray fluence only corresponds to 
a small fraction of the total mass of the star. One can still apply 
Eq.(\ref{Eq_Mdot}) to estimate the accretion rate, except that one 
should replace $M_*$ by $\zeta M_*$, where $\zeta < 1$ is the fraction
of stellar mass that is accreted after May 28. Noticing $z=1.1853$
(Cenko et al. 2011), this gives a peak accretion rate
\begin{equation}
\dot{M}_{\rm peak} ({\rm J12058}) \simeq 1.72 \times 10^{-7} \zeta_{-1}M_* 
{\rm s}^{-1}~,
\end{equation}
where $\zeta = 0.1 \zeta_{-1}$ has been adopted.

The constraint on BH spin for Sw J2058+05 is shown in Fig.\ref{fig_Ma}b. 
We find that the demand for BH spin is even more stringent for this source.
For $M_* = 1 M_\odot$, the required range of $a_{\bullet}$ is $(0.49, 0.98)$, with 
the most probably value $a_{\bullet} = 0.89$. For $M_* = 0.1 M_\odot$ (more probable), 
the range of $a_{\bullet}$ is $(0.81, 0.998)$ with the most probable value 
$a_{\bullet} = 0.99$.

\section{Conclusion and Discussion}

Sw J1644+57 and Sw J2058+05 are the first two proto-type objects in the
newly identified astrophysical phenomenon, namely, a relativistic jet
associated with a TDE from a supermassive BH. A straightforward 
question is why only some TDEs launch jets. Based on observational
properties, it has been argued that the jet has to be Poynting-flux
dominated (Burrows et al. 2011; Shao et al. 2011). Invoking the BZ
mechanism as the power of the jet, we show here that both events need
to invoke a BH with a moderate to rapid spin in order to interpret
the observations: the most probably values are $a_{\bullet} ({\rm J1644})
= 0.63, 0.90$ and $a_{\bullet} ({\rm J2058}) = 0.89, 0.99$ for
$M_* = 1, 0.1 M_\odot$, respectively. We therefore suggest that BH spin 
is the key factor behind the Sw J1944+57 - like events, although other 
factors may also play a role (e.g. Cannizzo et al. 2011).

An elegant feature of the method we employ is that the inferred BH
spin parameter $a_{\bullet}$ essentially does not depend on the BH mass.
On the other hand, knowing $a_{\bullet}$ leads to better constraint on BH 
mass based on the variability argument. For example, the observed
minimum variability time scale of J1644+57 is $\delta t_{\rm obs,min}
\sim 100$ s. If one relates $\delta t_{\rm min} = \delta t_{\rm obs,min}
/(1+z)$ to the time scale defined by the innermost radius of the accretion
disk, $r_{\rm in}/c$, one can derive the BH mass of Sw J1644+57 
\begin{equation}
M_{\bullet,6} \simeq 15 \left(\frac{r_{\rm in}}{r_g}\right)^{-1} 
\left(\frac{\delta t_{\rm obs,min}}{100s}\right)~.
\end{equation}
One has $2.5<M_6<15$ for $0 \leq a_{\bullet} \leq 1$, with the most
probably value $M_6 = 6.5$ for $a_\bullet = 0.9$. This is consistent 
with the constained BH mass from the $M - L_{\rm bulge}$ relation, which 
gives an upper limit of $2 \times 10^7 M_{\sun}$ (Burrows et al. 2011).

From Eq.(\ref{eq_Lmag}), one can also infer the strength of the 
magnetic field at the BH horizon
\begin{equation}
B_{\bullet,6} \simeq 131 f_b^{1/2} \eta^{-1/2} F(a_{\bullet})^{-1/2} a_{\bullet}^{-1} M_6^{-1}.
\end{equation}
Taking BH mass $M_{\bullet}=6.5\times 10^6 M_{\sun}$ and the most probaly 
value for BH spin $a_{\bullet}=0.9$,  one finds that the magnetic field 
threading BH would be $B_{\bullet} \sim 1.1 \times 10^6 {\rm G}$, which is much 
higher than the average field strength of a typical main sequence star 
($< 10^3 {\rm G}$). The accumulation of magnetic flux by accretion flow and 
instability in the disk may account for such high magnetic field stength 
(e.g. Tchekhovskoy et al. 2011). 

For Sw J2058+05, due to the low X-ray flux at the late epochs a much 
looser constraint on variability $\delta t_{\rm obs,min} < 10^4$ s was 
obtained (Cenko et al. 2011), so that the precise values of $M_{\bullet}$ 
and $B_\bullet$ cannot be derived.

An alternative scenario to interpret Sw J1644+57 - like event may be
the onset of an AGN (Burrows et al. 2011). This scenario, which predicts
that the two sources will be active at least in the following millennium, 
may be less favored in view of the rapid onset of emission in Sw J1644+57
and the gradual decay in both events, but is not ruled out.
Our method can be applied to this scenario as well, except that $f_b$
becomes much larger (due to the intrinsic rarity of AGN onset events),
and $M_*$ is no longer limited to the range of $0.1-10 M_\odot$. Our
analysis suggests that the BZ power is likely not adequate to interpret
the data (because of the large emission power demanded by the small
$f_b$ factor) even for maximum spin ($a_{\bullet} \sim 1$), unless the 
accretion rate is much higher, so that the total amount of fuel $M_* \gg 1
M_\odot$. This is not impossible since the fuel on the AGN onset scenario 
is from a gas cloud near the BH, whose mass is not specified.

Krolik \& Piran (2011) proposed a model for Sw J1644+57 invoking a white 
dwarf being tidally disrupted by a smaller BH ($M_\bullet \sim 10^4 M_\odot$). 
Regardless of how this model may interpret the $\delta t_{\rm min}$ and
the apparent association of the source with the center of host galaxy, 
the derived $a_\bullet$ range also applies to their model for the $M_*$ 
range of a white dwarf, since our constraint is $M_\bullet$-independent.

Finally, in our calculations we did not consider evolution of $a_\bullet$ 
during the accretion phase. This is justified given the large 
$M_\bullet / M_*$ ratio.

\acknowledgements This work is supported by NSF under Grant No. AST-0908362, 
by NASA under Grant No. NNX10AD48G, by National Natural Science 
Foundation of China under Grant No. 11003004, and National Basic Research 
Program (``973'' Program) of China under Grant No. 2009CB824800.
WHL acknowledges a Fellowship from China Scholarship Program for support.


\clearpage

\end{document}